# Plasmonic structure integrated single-photon detectors optimized to maximize polarization contrast

M. Csete[1]*, A. Szenes[1], D. Maráczi[1], B. Bánhelyi[2], T. Csendes[2] and G. Szabó[1]

[1]Department of Optics and Quantum Electronics, University of Szeged, H-6720 Szeged, Dóm tér 9, Hungary

[2]Department of Computational Optimization, University of Szeged, H-6720 Szeged, Árpád tér 2, Hungary

*Mária Csete: mcsete@physx.u-szeged.hu

**Abstract**: Numerical optimization was performed via COMSOL Multiphysics to maximize the polarization contrast of superconducting nanowire single photon detectors (SNSPDs). SNSPDs were integrated with four different types of one-dimensional periodic plasmonic structures capable of mediating p-polarized photon selectivity to the niobium-nitride superconducting nanowire pattern. Optimization with two different criteria regarding the maximal tilting resulted in wavelength-scaled periodic integrated structures, which have different geometrical parameters, and exhibit different polar angle dependent optical response and dispersion characteristics, as well as accompanying near-field phenomena at the extrema. Polarization contrast of $6.37 \cdot 10^2$ and $3.28 \cdot 10^2$ / $6.86 \cdot 10^{11}$ and $1.42 \cdot 10^{11}$ / $1.81 \cdot 10^{13}$ and $7.87 \cdot 10^{12}$ / $1.90 \cdot 10^3$ and $1.15 \cdot 10^5$ can be achieved in nanocavity- / nanocavity-deflector- / nanocavity-double-deflector / nanocavity-trench-array integrated P-SNSPDs optimized with 85° and 80° criterion regarding the maximal tilting.

## Introduction

The near-field enhancement in proximity of plasmonic antennas results in enhanced light-matter interaction, which can be exploited in photodetection, a unique application shows that it makes possible to read out nanoscale bar-codes [1]. There are several examples of plasmonically improved photodetectors in the scientific literature. This is due to that a polarized light incident on a diffraction grating can be coupled into surface plasmon polaritons (SPP), moreover can be completely absorbed in specific configurations [2]. Concentric grating-coupler made possible to achieve 20-fold enhancement in Si-based photodetectors along with the improvement of the electrical properties [3]. The lateral and in depth light confinement achievable via slits was also utilized in case of Si detectors [4]. The combination of a slit and a surrounding grating promoted to achieve 250-fold enhancement in the absorption of the embedded semiconducting material [5]. Various types of plasmonic waveguide based detectors have been also developed [6-8].

Although, the polarization selectivity is a unique and significant property of plasmonic patterns, there are only a few examples in the literature, which describe efforts to improve and utilize these capabilities. C-shaped apertures integrated onto Ge detectors were capable of enhancing detection efficiency significantly and ensuring polarization sensitivity throughout a wide wavelength region [9]. A nanoscaled photodetector was created by embedding Ge into the arms of a sleeve dipole antenna, and a polarization contrast of ~20 was achieved [10]. The spectral and polarization selectivity of complex patterns was used to design plasmonic photon sorters [11].

Among photodetectors, devices possessing single-photon sensitivity are particularly important in quantum information processing (QIP). One of the most widely used photodetectors is the superconducting nanowire single-photon detector (SNSPD) [12-14]. The most important part of conventional SNSPDs is a superconducting wire made of e.g. niobium-nitride (NbN) in a meandered pattern on silica or sapphire substrate. These wires are usually 4 nm thick with a 2 nm oxide cover-layer on the top of them. Due to the absorption of an infrared (1550 nm) single-photon the superconducting state breaks in the nanowire, which results in a detectable change of the voltage signal in the readout electronics. The width of the wire has to be smaller than 90-110 nm to ensure optimal detection probability. This explains, why a long meandered NbN pattern is needed to obtain large absorption across the active area of the detector. However, the length of the NbN wires limits the recovery time of an SNSPD, and hence the speed of the detector. As a consequence, different nanostructures integrated into the active area of the detector, which can enhance the absorption of the NbN wires, are in great demand. Such a structure could maintain high absorptance and faster detection simultaneously, even in case of a relatively small NbN filling-factor. Further important properties of SNSPDs can be tailored by integrated structures as well, namely it was shown that the external dark count rate can be reduced by a multilayer bandpass filter [15].

In recent SNSPD research the primary purpose is to ensure high detection efficiency, while in specific polarization-coding quantum key distribution based QIP applications the polarization selectivity is also important [16]. It has been already shown that the polarization selectivity differs for bare and plasmonic structure integrated devices [17-19]. Significant p-polarization specific absorptance enhancement has been achieved via integrated plasmonic gratings [20-24]. By adjusting the width and pitch of superconducting wires, polarization extinction ratio of 22 was achieved, however with a very low detection efficiency of 12% [25]. High and polarization independent efficiency has been ensured via spiral absorbing patterns [26, 27], crossed gratings [28], and via high-index dielectric material based compensation method without symmetry improvements [29].

In our present study the purpose was to determine the optimal one-dimensional periodic plasmonic pattern integrated P-SNSPD configurations capable of maximizing the polarization contrast. Comparative study of four different types of integrated P-SNSPDs optimized by setting the criterion regarding the maximal polar angle to 85° and 80° was performed. These results help the users to consider, which initial criterion promotes better SNSPD performance in specific applications.

## **Methods**

In this study four different periodic plasmonic structure integrated NbN patterns with approximately one-wavelength periodicity ($p$) were inspected, since these result in a Rayleigh phenomenon and have the potential to enhance the absorptance of the superconducting NbN nanowires.

In *nano-cavity-array-integrated* (NCAI)-SNSPD $l_{deflector}$ = 50-500 nm long gold segments were aligned along the absorbing NbN wires, by filling up the space between them completely (Fig. 1ba, ca). These gold segments were closed by a $t$ =50-70 nm thick gold layer, nominated as a reflector. All these gold segments together create an array of $w_{cavity}$ = 90-100 nm wide and $l_{cavity}$ = 56-506 nm high MIM nano-cavities just above the NbN wire segments, which cavities are filled with hydrogen silsesquioxane (HSQ). In *nano-cavity-deflector-array-integrated* (NCDAI)-SNSPD additional $l_{deflector}$ = 50-500 nm long and $w_{deflector}$ [10 nm, $p$ - $w_{cavity}$ -5 nm ] wide lateral gold segments were integrated into the NbN pattern at the anterior side of the nano-cavities at the substrate interface, which are referred as anterior deflectors and form a secondary $p$ -scaled array (Fig. 2ba, ca). In *nano-cavity-double-deflector-array-integrated* (NCDDAI)-SNSPD additional $l_{deflector}$ = 50-500 nm long and $w_{deflector}$ [10 nm, $p$ - $w_{cavity}$ -20 nm] wide gold deflectors were integrated into the NbN pattern both at the anterior and the exterior sides of nano-cavities at the substrate interface (Fig. 3ba, ca). The length and width of the two deflectors have the same geometric bounds, however these parameters were tuned independently. In *nano-cavity-trench-array-integrated* (NCTAI)-SNSPD the NbN wires were surrounded by $l_{deflector}$ = 50-500 nm long and $w_{deflector}$ [10 nm, $p$ - $w_{cavity}$ -20 nm] wide gold segments laterally, which formed a secondary nano-cavity-array between the NbN wires (Fig. 4ba, ca). In NCDDAI-P-SNSPD and NCTAI-P-SNSPD optimization, the anterior and exterior deflectors contact was prevented by setting a constraint of 10 nm regarding their minimal distance.

The above described SNSPDs were modeled in RF module of COMSOL Multiphysics. This software solves the Maxwell equations numerically and allows to obtain the optical response of an integrated periodic structure with an arbitrary geometry. The incoming light was a 1550 nm wavelength, p- or s-polarized plane wave defined via a port below the NbN segments, which corresponds to the substrate side illumination. During simulations only one single unit cell of the periodic pattern was modeled, and the phase matching between unit cells was ensured via Floquet periodic condition at the vertical boundaries. The reflection and transmission of the inspected SNSPD design was obtained by integrating the power flow through imaginary horizontal interfaces at the reflection and transmission side of the detector, respectively. The absorptance of NbN segments was determined based on the Joule-heating, as it is described in our previous paper [18]. The ratio of absorptance of p- and s-polarized light is referred as polarization contrast throughout this paper.

To find the optimal configuration capable of maximizing the polarization contrast the geometry and illumination direction of different plasmonic structure integrated devices has been optimized. Namely, the $w_{cavity}$ width and $l_{cavity}$ length of nano-cavities, $w_{deflector}$ and $l_{deflector}$ width and length of deflectors, the $t$ thickness of the reflector and the $w = w_{cavity}$ width of NbN wires, as well as the p periodicity in ~ $\lambda_{SPP}$: [1000 nm, 1100 nm] region were all subjects of optimization. The illumination direction was also tuned, the $\varphi$ angle between the detector surface normal and the illumination direction is referred as the polar angle, similarly to our previous studies [18, 19, 21-24]. The azimuthal angle was fixed to $\gamma$ =90°, in this orientation the **E**-field of p-polarized light oscillates perpendicularly to the integrated plasmonic pattern. The selection of S-orientation is explained with that this orientation results in maximal p-polarized absorptance in A-SNSPDs optimized for absorptance, according to our previous studies [18, 19, 21-24].

To perform the optimization, the in-house developed GLOBAL optimization methodology was implemented using LiveLink for MATLAB in RF module of COMSOL Multiphysics [30, 31]. To analyze the polarization contrast, which was used in the objective function of the optimization, two models were evaluated via LiveLink for Matlab. The optimizations have been performed by setting a criterion of $\varphi$ = 85° and 80° regarding the maximal tilting to design devices for operation below these maximal tilting. The comparative study of optimized devices nominated as P-85-SNSPD and P-80-SNSPD was realized. For all optimized systems the polar angle dependent optical responses at 1550 nm and the dispersion characteristics in [0°, 88°] polar angle and [600 nm, 2500 nm] wavelength interval were determined in S-orientation. The locations corresponding to the largest values of absorptance and polarization contrast taken on in the inspected 0°-80°/85° polar angle interval are nominated as global maxima, even if the signals exhibit further increasing characteristics. Finally, the near-field distribution was inspected at the extrema of polarization contrast to uncover the underlying nanophotonical phenomena. All s-polarized absorptance signals are multiplied by $10^{13}$ to ensure comparability.

## Results

### *Comparative study of NCAI-P-SNSPDs*

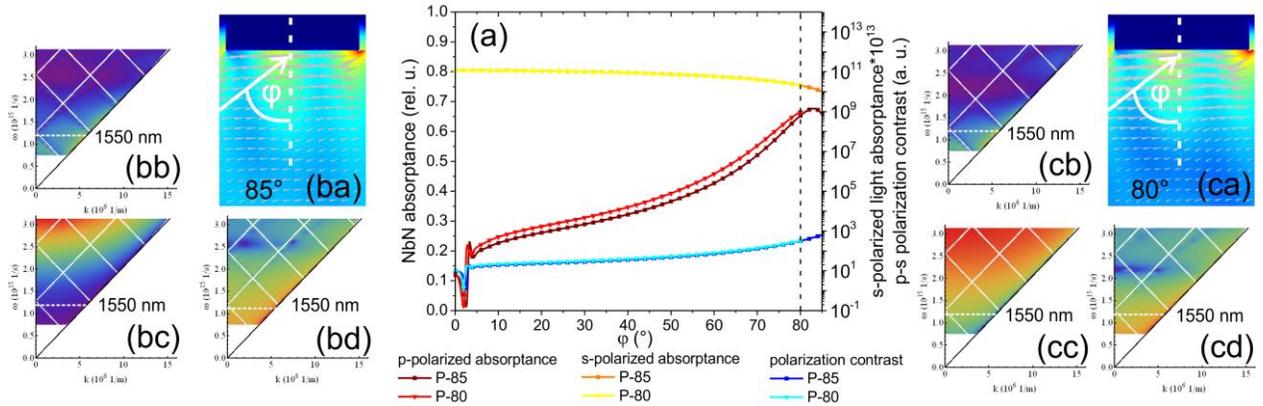

*Fig. 1.* (a) Polar angle dependent p- and s-polarized light absorptance and polarization contrast of NCAI-P-85-SNSPD and NCAI-P-80-SNSPD. (ba, ca) Near-field distribution, dispersion diagrams in (bb, cb) p-polarized and (bc, cc) s-polarized light absorptance and (bd, cd) polarization contrast of (b) NCAI-P-85-SNSPD and (c) NCAI-P-80-SNSPD.

In optimized NCAI-P-SNSPDs the course of the polar angle dependent absorptance is very similar (Fig. 1a, STable 1. in supplementary data). Both NCAI-P-85 and NCAI-P-80 systems show low absorptance at small polar angles, which rapidly increases with tilting. The collective resonance on the integrated nano-cavity-array is at play throughout the complete polar angle interval, while at small tilting an intense optical response modulation is also observable. The global minimum-maximum pairs indicate -1 order grating-coupling on the wavelength-scaled nano-cavity-array into plasmonic and photonic modes, respectively (Fig. 1a). The absorptance is the highest near the plasmonic Brewster angle (PBA) of the integrated pattern in both optimized NCAI-P-SNSPDs. The PBA is the polar angle, where the transmittance / reflectance on the nanocavity-array reaches its maximum / minimum due to impedance matching, and the absorptance in cavity loadings reaches its maximum. The PBA can be calculated as:

$$\cos\varphi_{PBA} = (k_{surfacewave} \cdot w)/(k_{photon} \cdot p) \qquad (1)$$

where $w$ is the width of the nano-cavities, $p$ is the period of the integrated pattern, $k_{surfacewave}/k_{photon}$ is the wave vector of coupled surface wave / incoming photon [32, 33].

The 67.65% global p-polarized absorptance maximum of NCAI-P-85-SNSPD is slightly larger than the 66.57% absorptance maximum achieved in NCAI-P-80-SNSPD (Fig. 1a, STable 1). The $\varphi_{abs\_max}^{NCAI-P-85}$=83° and $\varphi_{abs\_max}^{NCAI-P-80}$=80° polar angles corresponding to the global absorptance maxima are nearby the PBA of the integrated systems, which appear at $\varphi_{PBA}^{NCAI-P-85}$=84.97° and $\varphi_{PBA}^{NCAI-P-80}$=85.02°, respectively. This indicates that the optimization resulted in geometries, which are capable of enhancing absorptance via PBA phenomenon.

The geometric parameters are approximately the same, which explains the similar optical responses, the only significant difference is between the cavity lengths, which promotes larger NbN/Au ratio in the optimized NCAI-P-80 (STable 1). The cavity width both in optimized NCAI-P-85 and NCAI-P-80 is at the lower bound, namely it is $w_{cavity}$=90. The larger $l_{cavity}/(\lambda/4)$ ratio, , in NCAI-P-85 indicates less squeezed MIM cavity modes, and operation closer to a quarter-wavelength cavity. The NCAI-P-80 optimization resulted in larger NbN/Au volume ratio, which can explain, that the absorptance is higher at the same polar angle almost throughout the complete inspected interval. Interestingly, a nano-cavity shorter than a quarter-wavelength results in a slightly larger p-polarized absorptance in NCAI-P-80, indicating that the material volume fraction determines the achieved absorptance, rather than the local effects originating from cavity resonance.

The course of the polar angle dependent polarization contrast is governed by the modulations observable in p-polarized absorptance, namely minimum-maximum pairs appear also on both polarization contrast signals (Fig. 1a). The achieved polarization contrast values are determined by the rapidly decreasing s-polarized absorptance. The contrast is higher at the same polar angles throughout 80° in NCAI-P-80, where it exhibits the 3.28·10$^2$ global maximum. However, the highest 6.37·10$^2$ contrast is reached at 85° in the optimized NCAI-P-85 system, due to the more rapidly decreasing s-polarized absorptance at larger tilting.

In the optimized NCAI-P-85 at the 85° tilting corresponding to the polarization contrast maximum the absorptance is close to its global maximum. In optimized NCAI-P-80 the 80° location of the absorptance and contrast maximum are coincident, and the achieved absorptance is slightly larger than the absorptance reached at 85° in optimized NCAI-P-85. In contrast, the polarization contrast is approximately two-times smaller. This proves that the course of the s-polarized absorptance strongly influence the polarization contrast, and can result in reversal ratio in the achieved polarization contrast values. Both in NCAI-P-85 and in NCAI-P-80 devices the contrast remains higher than 10 throughout the complete inspected interval excluding the ~5° wide polar angle region of significant optical response modulations, which are caused by grating couplings.

Based on the dispersion diagrams of NCAI-P-SNSPDs, the PBA phenomenon results in a global absorptance maximum on the p-polarized absorptance at the boundary of the second Brillouin zone

(Figure 1bb and cb). The global-local minimum-maximum pairs observed on the polar angle dependent NbN absorptance at 1550 nm are located on the left-tilted branches corresponding to the -1 order grating-coupling into plasmonic and photonic modes on the integrated periodic patterns (Figure 1bb, cb). The high polarization contrast is promoted by the s-polarized light absorptance, which is depressed inside a wide frequency and polar angle band at the boundary of second Brillouin zone (Figure 1bc, bd, cc, cd).

On the near-field distribution of NCAI-P-85 and NCAI-P-80 one can see that the resonant metal-insulator-metal (MIM) modes are slightly squeezed (Fig. 1ba, ca). Significant **E**-field enhancement is observable around the NbN stripes at the anterior side of the loaded nano-cavities, which are slightly shorter than quarter-wavelength. Caused by the large polar angle corresponding to the PBA, the Poynting vector is almost parallel to the interleaved gold segments in both optimized NCAI-P-SNSPDs. The time-evolution of the **E**-field shows that the neighboring nano-cavities are intermittently shined by forward propagating waves (see the supplementary video Media 1 and Media 2).

***Comparative study of NCDAI-P-SNSPDs***

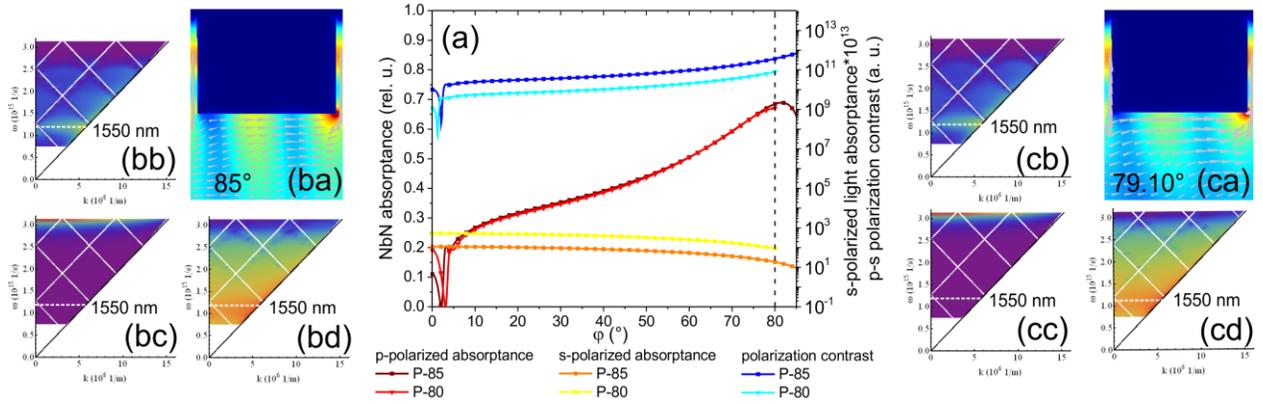

**Fig. 2.** (a) Polar angle dependent p- and s-polarized light absorptance and polarization contrast of NCDAI-P-85-SNSPD and NCDAI-P-80-SNSPD. (ba, ca) Near-field distribution, dispersion diagrams of (bb, cb) p-polarized and (bc, cc) s-polarized light absorptance and (bd, cd) polarization contrast of (b) NCDAI-P-85-SNSPD and (c) NCDAI-P-80-SNSPDs.

Similarly to the NCAI-SNSPDs, the polar angle dependent p-polarized absorptance of NCDAI-P-85 and NCDAI-P-80 is almost equal. A slight difference can be observed in their course only at small polar angles, in the regions of local extrema caused by grating-coupling (Fig. 2a, STable 2). According to the grating-coupling condition of

$$\lambda_{surfacewave} = \frac{1}{\frac{n_{silica}}{\lambda_{photon}} \cdot sin\,\varphi - \frac{1}{p}} \quad (2)$$

where $\lambda_{surfacewave}$ refers to the $\lambda_{photonic}$= 1068.75 nm or $\lambda_{SPP}$ = 1057.12 nm wavelength of the photonic or plasmonic mode wavelength, respectively, $\lambda_{photon}$= 1550 nm is the incident wavelength and $n_{silica}$ is the refractive index of silica substrate, in case of a larger period a smaller incidence angle makes possible coupling into a specific backward propagating surface mode. Both NCDAI-P-SNSPD couples in -1 order into a plasmonic mode, and a global minimum appears at smaller 2.1° and larger 3.1° tilting in NCDAI-P-85 and NCDAI-P-80, according to their larger 1022.03 nm and smaller 1006.00 nm period, respectively. After a monotonous increase with increasing polar angle, the 68.93% and 68.81% absorptance maxima are reached at $\varphi_{abs\_max}^{NCDAI-P-85}$ =81.6° and $\varphi_{abs\_max}^{NCDAI-P-80}$ =80° polar angles. The former is close to the $\varphi_{PBA}^{NCDAI-P-85}$ =84.95° PBA of NCDAI-P-85, while in NCDAI-P-80 the $\varphi_{PBA}^{NCDAI-P-80}$ =84.80° PBA more significantly differs from the latter. The relationship show, that again in case of NCDAI-P-85 the larger global maximum is ensured via PBA phenomenon, while in NCDAI-P-80 the p-polarized absorptance maximum is just slightly promoted by PBA phenomenon.

All geometrical parameters are similar, the slightly smaller period, cavity length and deflector dimensions in NCDAI-P-80 optimization indicate a less robust gold grating (STable 2). The deflector lengths approximate the upper bound of 500 nm in both cases, while they fill horizontally the space between the nano-cavities almost completely and form an extended cavity at the substrate interface. The approximately $2.5 \cdot \lambda/4$ long extended cavities reveal that the **E**-field concentration around the NbN segments is compromised in presence of deflectors. The absorptance is larger in NCDAI-P-85-SNSPD almost throughout the complete polar angle interval, in accordance with the larger NbN/Au volume fraction ratio.

The array of robust deflectors results in 9 order of magnitude larger polarization contrast compared to NCAI-P-SNSPDs. This due to their strong s-polarized light absorptance depressing effect, since the p-polarized absorptance is decreased by 1.14% and 0.27% at tilting corresponding to polarization contrast maxima with respect to the corresponding absorptances in NCAI-SNSPDS. Significant difference is that the NCDAI-P-85 shows five-times higher contrast in the entire polar angle interval, except in the interval of modulations originating from the polar angle dependent p-polarized absorptance (Fig. 2a). The superiority of P-85 is due to the more strongly depressed s-polarized light absorptance.

In optimized NCDAI-P-85 the maximal $6.86 \cdot 10^{11}$ polarization contrast at 85° is approximately five-times larger than the $1.42 \cdot 10^{11}$ contrast maximum in NCDAI-P-80 at 80°. In case of applications, which require extreme polarization contrast through a wide polar angle interval, NCDAI-P-85 is an ideal construction. Even though the NCDAI-P-80 shows by 1.82% larger absorption at 79.1° than NCDAI-P-85 at 85°, the polarization contrast is approximately five-times smaller. Important advantage of these devices is that in NCDAI-P-85 / NCDAI-P-80 the global polarization contrast minimum is still larger than $10^8$ / $10^7$.

Both NCDAI-P-SNSPDs show high p-polarized light absorptance at the boundary of the second Brillouin zone due to PBA phenomenon, similarly to NCAI-P-SNSPDs (Figure 2bb, cb). The global-local minimum-maximum pairs on the p-polarized absorptance at 1550 nm are located on the left-tilted branches corresponding to -1 order grating-coupling into plasmonic and photonic modes (Figure 2bb, cb). In case of NCDAI-P detectors a significantly higher polarization contrast is achievable due to the more depressed s-polarized light absorptance, which proves the polarization specific absorption enhancing role of deflectors (Fig. 2bc, bd, cc and cd).

In case of NCDAI-P-SNSPDs the deflectors' geometrical parameters reach the upper limit during the optimization process which indicates also, that high polarization contrast can be reached via robust deflectors resulting in appearance of extended MIM cavities (Figure 2ba, ca). The highest near-field enhancement is at the anterior side of the $\sim 2.5 \cdot \lambda/4$ long extended MIM cavity entrance again. Large enhancement is observable around the NbN segments at the entrance of the $\sim 0.5 \cdot \lambda/4$ long inner nano-cavities as well. The highest polarization contrast can be reached at large polar angles, close to the PBA of NCDAI-P-SNSPDs. Accordingly, the near-field distribution indicates Poynting vectors nearly parallel to the substrate interface of wide deflectors (Fig. 2ba, ca). Below the nanocavity entrances the power-flow points towards the NbN stripes inside the extended MIM cavities. The time-evolution of the **E**-field indicates that the neighboring cavities are shined cyclically and synchronously in NCDAI-P-85, while in NCDAI-P-80 their illumination is asynchronous (see the supplementary video Media 3 and Media 4).

### *Comparative study of NCDDAI-P-SNSPDs*

In case of nano-cavity-double-deflector integrated NCDDAI-P-SNSPDs the course of the polar angle dependent p-polarized absorptances is very similar (Fig. 3, STable 2). The integrated structure couples into photonic modes at the global minima of 1.4° and 1.3° in +1 and -1 order in case of NCDDAI-P-85 and NCDDAI-P-80, respectively. Significant difference is that a local minimum is observable at 71.9°, where the NCDDAI-P-85 couples into a photonic mode in -2 order. The PBA of NCDDAI-P-85 and NCDDAI-P-80 are $\varphi_{PBA}^{NCDDAI-P-85}$=85.33° and $\varphi_{PBA}^{NCDDAI-P-80}$=85.11° according to equation (1).

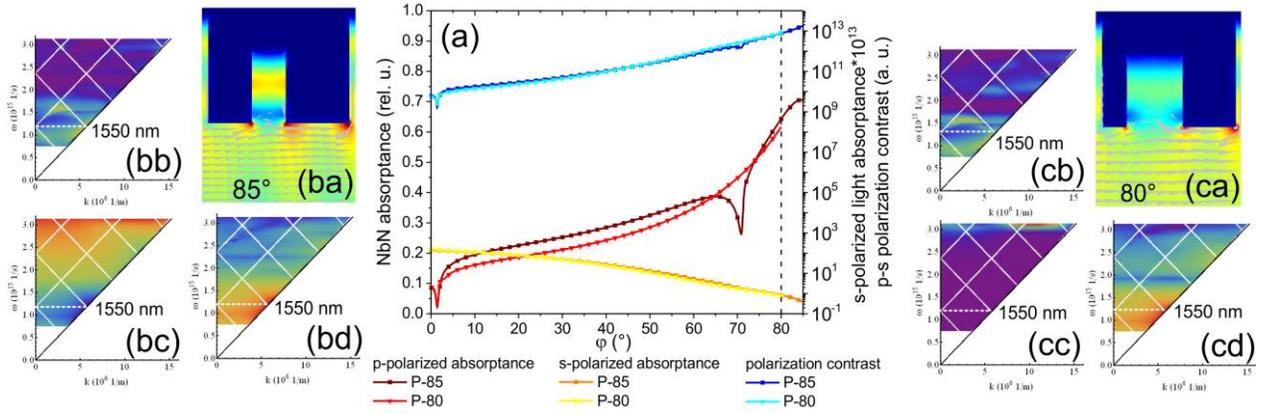

**Fig. 3.** (a) Polar angle dependent p- and s-polarized light absorptance and polarization contrast of NCDDAI-P-85-SNSPD and NCDDAI-P-80-SNSPD. (ba, ca) Near-field distribution, dispersion diagrams of (bb, cb) p-polarized and (bc, cc) s-polarized light absorptance and (bd, cd) polarization contrast of (b) NCDDAI-P-85-SNSPD and (c) NCDDAI-P-80-SNSPDs.

These PBA values are approximated by the observed $\varphi_{abs\_max}^{NCDDAI-P-85}$=84.4° and $\varphi_{abs\_max}^{NCDDAI-P-80}$=80° tilting corresponding to global absorptance maxima, however better agreement is found in case of NCDDAI-P-85 (Fig. 3a). The largest 70.58% absorptance maximum is reached in the optimized NCDDAI-P-85, while the NCDDAI-P-80 exhibits significantly smaller 61.66% global absorptance maximum. Moreover, NCDDAI-P-85 makes possible to reach a larger p-polarized light absorptance almost throughout the complete polar angle dependent interval, except the regions of grating-couplings.

The inner nano-cavity length is slightly smaller than $\lambda/4$ in NCDDAI-P-SNSPDs, both cavity widths are at the lower bounds, as in case of the previous optimizations, while the deflector lengths are at the upper bound uniformly (STable 2). In both cases the anterior deflector width is significantly larger than the exterior one. In both NCDDAI-P-SNSPDs the extended cavities are approximately $3 \cdot \lambda/4$ long, which reveals that the **E**-field concentration around the NbN stripes is better than in NCDAI-P-SNSPDs. The NbN/Au volume fraction ratio is larger in NCDDAI-P-85 than in NCDDAI-P-80, which is in accordance with the larger absorptance.

By comparing NCDDAI- with NCDAI-SNSPDs, based on the larger NbN fill-factor and better approximation of the $3 \cdot \lambda/4$ length one can conclude that larger absorptance can be reached in NCDDAI-SNSPDs. However, in NCDDAI-P-80 the absorptance is smaller than in NCDAI-P-80, which indicates that the synchronization of the propagating modes, and the confinement of the localized modes together determine the achieved absorptance.

The polar angle dependent polarization contrast of NCDDAI-P-SNSPDs follows the course of the polar angle dependent absorptances, as expected from the previous cases, namely the local minimum on the contrast is caused by the -2 order grating coupling in NCDDAI-P-85 (Fig. 3a). Important to note, that NCDDAI-P-85 shows the highest, $1.81 \cdot 10^{13}$ polarization contrast among all of the optimized SNSPDs, while NCDDAI-P-80 allows to reach the second largest $7.87 \cdot 10^{12}$ polarization contrast. At 85° tilting of NCDDAI-P-85 the polarization contrast is larger than $10^{13}$ due to the strongly depressed s-polarized light absorptance. The contrast monotonously increases from smaller polar angles, but it remains larger than $10^{9}$ throughout the complete polar angle interval. The polarization contrast of optimized NCDDAI-P-80 shows the same characteristics, however it is smaller / larger at smaller / larger polar angles.

By comparing NCDDAI- with NCDAI-P-SNSPDs, one can conclude that 26-fold and 55-fold larger polarization contrast is achievable in NCDDAI-P-SNSPDs despite of the larger extended cavity length and the presence of double-deflector array. The absorptance at the contrast maximum is increased by 5.87% and decreased by 4.64% with respect to corresponding maximum NCDAI-Ps, which proves that the relative polarization contrast enhancement in NCDDAI-Ps is due to the more suppressed s-polarized absorptance (Fig. 3a).

The highest p-polarized absorptances are achievable at the boundary of the second Brillouin zones, similarly to the previous cases (Fig. 3bb, cb). The global minimum at 1550 nm in the optimized NCDDAI-P-85 / 80 is located on the right / left tilted branch corresponding to +1/-1 order coupling into photonic modes (Fig 3a). The local minimum observed in case of NCDDAI-P-85 is located on a left tilted photonic branch originating from -2 order grating-coupling (Fig. 3a, bb). The s-polarized light absorptance is strongly depressed inside wide bands due to the polarization selectivity of deflectors, which results in a strong polarization contrast inside analogously wide bands (Fig. 3bc, bd, cc, cd).

The near-field of NCDDAI-P-SNSPDs is very similar due to the similar geometric parameters found during optimization (Fig. 3ba, ca). The most significant **E**-field enhancement occurs at the anterior side of nano-cavity entrances. Relatively smaller enhancement is observable inside the cavities consisting of two anti-nodes, between the deflectors and at the corners of the intersecting cavities as well. The power-flow is the strongest laterally in the x direction below the deflectors, while considerable amount of energy flows towards the nano-cavities inserted into the deflector arrays as well. The nano-cavities are shined cyclically, and synchronously, but with more intense squeezed MIM modes in NCDDAI-P-85 (see the supplementary video Media 5 and Media 6).

### *Comparative study of NCTAI-P-SNSPDS*

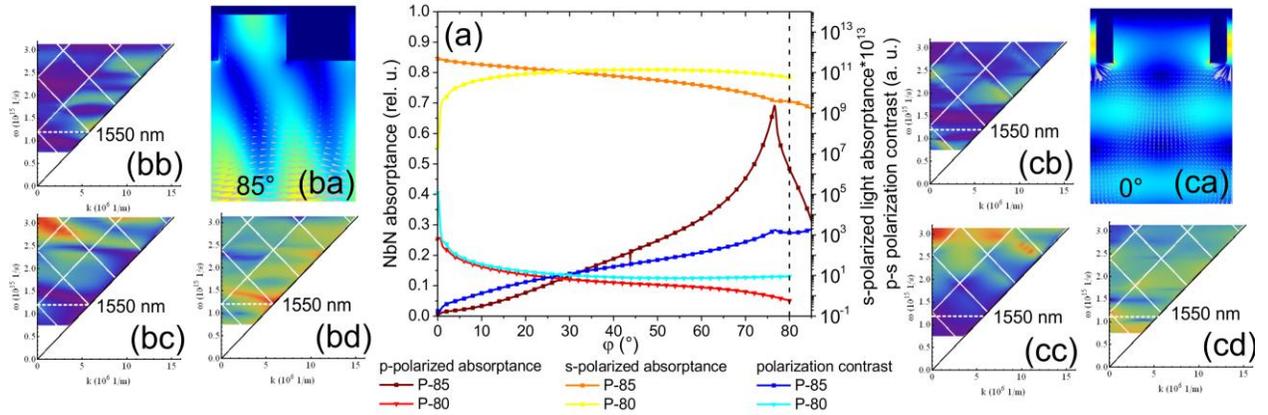

**Fig. 4.** (a) Polar angle dependent p- and s-polarized light absorptance and polarization contrast of NCTAI-P-85-SNSPD and NCTAI-P-80-SNSPD. (ba, ca) Near-field distribution, dispersion diagrams of (bb, cb) p-polarized and (bc, cc) s-polarized light absorptance and (bd, cd) polarization contrast of (b) NCTAI-P-85-SNSPD and (c) NCTAI-P-80-SNSPDs.

The optimized NCTAI-P-85 shows very small p-polarized absorptance at perpendicular incidence, which begin to increase throughout larger polar angles, as in case of the previous optimized P-SNSPDS, and reaches its 69.08% absorptance maximum at $\varphi_{abs\_max}^{NCTAI-P-85}$ =76.6°, where the integrated pattern couples the incoming light into photonic modes in -2 order (Fig. 4a, STable 1). This example proves that in NCTAI-P-SNSPD the same grating-coupling is capable of resulting in global maximum at tilting, where global minimum appears in NCDDAI-P-85, due to the absence of into-substrate-deepened deflectors. The PBA of NCTAI-P-85 and NCTAI-P-80 is $\varphi_{PBA}^{NCTAI-P-85}$ =85.27° and $\varphi_{PBA}^{NCTAI-P-85}$ =85.19° according to equation (1), however NCTAI-SNSPDs do not show global or local absorptance maxima in proximity of these polar angles. In contrast, in the optimized NCTAI-P-80 the optical response differs significantly from that in all previous cases. The 25.69% global p-polarized absorptance maximum is reached at $\varphi_{abs\_max}^{NCTAI-P-80}$ =0.5°, where the absorptance starts to decrease towards larger polar angles. In NCTAI-P-80 at the global p-polarized absorptance maximum the incoming light couples simultaneously into plasmonic / photonic modes in +1 / -1 order, respectively. These results indicate, that in contrast to the previous cases, in NCTAI-P-SNSPDs the global absorptance maxima are ensured via near-field enhancement originating from synchronized surface modes.

The periods of the meandered NbN stripe patterns and the cavity widths are very similar in NCTAI-P-SNSPDs, however the cavity lengths and deflector geometries differ significantly (STable 1), which results in considerably different optical responses. In NCTAI-P-85 the deflector geometry is strongly asymmetric (Fig. 4ba), the anterior deflector width is much larger than the exterior one, as in the NCDDAI-P-SNSPDs. The exterior deflector width is 10 nm narrow, which corresponds to the lower bound. In contrast, NCTAI-P-80 optimization shows an almost symmetric interleaved deflector geometry (Fig. 4ca), consisting of relatively narrow, but almost identical 127.27 nm and 126.7 nm deflectors. In the optimized NCTAI-P-85 the cavities are capable to support $\lambda/2$ cavity modes while in NCTAI-P-80 the cavities are approximately $\lambda/4$ long. The ratio of absorptances is reversal with respect to the NbN/Au volume fraction ratio (STable 1). This indicates that the **E**-field enhancement originating from coupled propagating modes can overcompensate the non-perfect overlap of the localized mode and the NbN wires, and the smaller volume fraction of NbN in NCTAI-P-85.

The larger $1.15 \cdot 10^5$ polarization contrast can be reached via NCTAI-P-80 optimization at 0.5°, which tilting is unique among all the inspected SNSPDs (Fig. 4a). The polarization contrast decreases towards larger tilting according to the characteristics of polar angle dependent p-polarized absorptance and reaches 10 at 80° angle of incidence. The optimized NCTAI-P-80 is the most appropriate for applications, which prefer perpendicularly incident light, but it is less efficient in other polar angle intervals.

In the optimized NCTAI-P-85 detector the polarization contrast is very small at perpendicular incidence, moreover indicates that s-polarized light is absorbed more effectively, while for larger polar angles the polarization contrast in NCTAI-P-85 overrides the contrast in NCTAI-P-80 (Fig. 4a). After the global maximum the p-polarized absorptance begins to decrease, however at 85° it is ~31.42%, while the contrast reaches the $1.9 \cdot 10^3$ global maximum. From the applications point of view it is preferable to operate at the global maximum of absorptance. This means only a slightly smaller $1.72 \cdot 10^3$ polarization contrast but a significantly larger ~69.08% absorptance.

The dispersion curve of optimized NCTAI-P-85 shows the highest p-polarized absorptance at the boundary of the second Brillouin zone again, while on the dispersion map of NCTAI-P-80 enhanced p-polarized absorption is observable due to the appearance of an inverted mini-gap at perpendicular incidence (Fig. 4bb, cb). The large global maximum on the p-polarized light absorptance of the optimized NCTAI-P-85 at 1550 nm is located on a left-tilted branch originating from -2 order grating-coupling into photonic modes (Fig. 4bb). In contrast, in the optimized NCTAI-P-80 the p-polarized light absorptance and contrast maxima appear inside the inverted minigap, which is opened in between branches of +1 / -1 order coupling into plasmonic / photonic modes (Fig. 4cb). In the absence of deflectors the s-polarized light absorptance is less depressed (Fig 4bc, cc), but it is still low and results in a relatively large polarization contrast inside a corresponding branch in NCTAI-P-85 and in inverted minigap in NCTAI-P-80, respectively (Fig 4bd, cd).

In optimized NCTAI-P-85 the significant near-field enhancement occurs in the substrate and inside the secondary cavity-array, which is nominated as trench-array (Fig. 4ba). The power mostly flows along the detector substrate interface, according to the large tilting corresponding to grating coupling. In case of NCTAI-P-80 the **E**-field enhancement is significant in the nano-cavities around the NbN segments, while noticeable enhancement is observable at the deflectors corners in the secondary cavity-array as well. The Poynting vectors point towards the nano-cavity entrances indicating the well-controlled energy flow towards the absorbing segments (Figure 4ca). The **E**-field enhancement occurs almost / completely synchronously in neighboring nano-cavities in NCTAI-P-85 / NCTAI-P-80, respectively (see the supplementary video Media 7 and Media 8).

## Conclusions

The geometry of four different types of plasmonic structure integrated SNSPDs was optimized to maximize the ratio of p- and s-polarized light absorptance by setting 85° and 80° polar angles as the maximal possible tilting. The polar angle dependent absorptance of the p-polarized light shows modulations at specific polar angles due to grating-coupling on the periodic integrated structures. The effects of PBA in p-polarized absorptance enhancement is dominant / noticeable in all of the inspected P-85 / P-80 SNSPDs, except the NCTAI-P-80-SNSPD. However, the achievable polarization contrast is determined by the s-polarized absorptance suppression. NCTAI-P-80-SNSPD is exceptional, which exclusively shows an exponentially decreasing characteristics in polarization contrast, mainly governed by the course of p-polarized absorptance. In all other systems the polar angle dependent polarization contrasts exponentially increase with increasing tilting, except those polar angles, where local modulations occur on the absorptance due to grating coupling.

The dispersion diagrams of the optimized P-SNSPDs show that the PBA phenomena appear in p-polarized absorptance at the boundary of second Brillouin zone, and the p-polarized global absorptance maxima at 1550 nm appear close to this in all cases except in NCTAI-P-80-SNSPD, where it is reached at the center of a minigap at perpendicular incidence. The p-polarized maxima are located inside bands of suppressed s-polarized absorptance at 1550 nm, which make possible to reach strongly enhanced polarization contrast. The highest polarization contrast is reached close to PBA at 1550 nm in both NCAI-/NCDAI-/NCDDAI-P-SNSPDs and in NCTAI-P-85-SNSPD, however in the latter the grating coupling of phase-synchronized modes promotes to reach the maximal absorptance at tilting smaller than the PBA.

We have shown evidence of coupled propagating and localized modes at the detector surface based on the EM-field distribution and time averaged power-flow. The coexistence of different coupled propagating modes promotes to reach simultaneously high absorptance and polarization contrast at approximately perpendicular incidence onto NCTAI-P-80-SNSPD.

The enhanced and the extremely large polarization contrasts in NCDAI- and NCDDAI-P-SNSPDs indicate the polarization selection role of deflectors. The polarization contrast exhibits correlation in all cases with the (extended) cavity length in quarter-wavelength units, additionally with the NbN/Au volume fraction except in NCAI-P-85 and with p-polarized absorptance, except in NCTAI-P-80, but correlates with the absorptance at the polarization contrast maximum only in NCDDAI-P-85.

## Acknowledgements

The research was supported by National Research, Development and Innovation Office-NKFIH through project "Optimized nanoplasmonics" K116362. Mária Csete acknowledges that the project was supported by the János Bolyai Research Scholarship of the Hungarian Academy of Sciences.